\documentclass[prf,amsmath,amssymb,reprint]{revtex4-1}
\usepackage{epsfig}                                                                 
\usepackage{dcolumn}                                                                
\usepackage{bm}                                                                     
\let\boldsymbol\pmb
\usepackage[pdfstartview=FitH,bookmarksopen=true,bookmarksopenlevel=1]{hyperref}    
\begin{document}
\title[Asymptotic reductions of the diffuse-interface model]%
{Asymptotic reductions of the diffuse-interface model,\\with applications to contact lines in fluids}
\author{E. S. Benilov}
 \altaffiliation[]{Department of Mathematics and Statistics, University of Limerick, Limerick, V94 T9PX, Ireland}
 \email{Eugene.Benilov@ul.ie}
 \homepage{https://staff.ul.ie/eugenebenilov/}
\date{\today}

\begin{abstract}
The diffuse-interface model (DIM) is a tool for studying interfacial dynamics.
In particular, it is used for modeling contact lines, i.e., curves where a
liquid, gas, and solid are in simultaneous contact. As well as all other
models of contact lines, the DIM implies an additional assumption: that the
flow near the liquid/gas interface is isothermal. In this work, this
assumption is checked for the four fluids for which all common models of
contact lines fail. It is shown that, for two of these fluids (including
water), the assumption of isothermality does not hold.

\end{abstract}
\maketitle

\section{Introduction}

The single most important open problem in hydrodynamics is that of contact
lines, i.e., curves where a liquid, gas, and solid are in simultaneous contact
(such as, for example, the circumference of a droplet on a substrate). It has
been known for almost fifty years \cite{HuhScriven71} that the Navier--Stokes
equations and the standard boundary conditions fail near a moving contact
line, yet there seems to be no consensus as to how this issue can be resolved
\cite{BonnEggersIndekeuMeunierRolley09,Velarde11}. The problem is caused by
the no-slip condition preventing the fluid particles on the contact line from
moving -- hence, the contact line itself is pinned to the substrate. As a
result, numerous phenomena involving wetting/dewetting (e.g., sliding
droplets) can be neither understood nor modeled.

Several attempts to remedy the problem have been made -- typically, by
modifying the boundary condition at the substrate in such a way that, near a
contact line, the fluid can slip (e.g., Refs.
\cite{HuhMason77,BenneyTimson80,Hocking1981,Gouin87,Shikhmurzaev93,Sharma93,Shikhmurzaev97,BenilovVynnycky13}%
). In some cases, different models agree with each other, in others they do
not \cite{SibleySavvaKalliadasis12,SibleyNoldSavvaKalliadasis14}. Furthermore,
it has been recently shown
\cite{PuthenveettilSenthilkumarHopfinger13,BenilovBenilov15} that there are
several fluids including water, for which none of the commonly used models
produces physically meaningful results\footnote{The only exception is the
interface-formation model proposed in Ref. \cite{Shikhmurzaev93} -- which,
however, involves 13 undetermined constants. These constants are specific to
each liquid/substrate combination and need to be pre-measured before the model
can be used.}.

Most importantly, all existing models of contact lines have one feature in
common: they assume that the flow near a liquid/vapor interface is isothermal.
In addition, most theories assume that the Reynolds number based on the
interfacial thickness is small. Yet, neither of these assumptions has been
verified. Direct measurements at such small scales are extremely difficult to
carry out, and nor can one draw conclusions about an interface from the
characteristics of the \emph{global} flow: even if the latter is isothermal,
the interface may not be.

Indeed, the high-gradient nature of the near-interface region can give rise to
strong production of heat due to viscosity and compressibility, as well as
evaporation and condensation. The released heat may cause strong, albeit
local, temperature variations, which can significantly affect the dynamics of
the contact line -- as can fluid inertia if the local Reynolds number is large.

In the present work, the so-called diffuse-interface model is used to check the
assumptions of isothermality and small Reynolds number for four fluids for
which the common models of contact lines fail. It is demonstrated that, for
water and mercury examined in Refs.
\cite{PodgorskiFlessellesLimat01,WinkelsPetersEvangelistaRiepenEtal11}, at
least one of the assumptions does not hold. For glycerol and ethylene glycol
examined in Ref. \cite{KimLeeKang02}, both assumptions actually hold -- hence,
the discrepancies between the experiments and theory in this case are due to
different reasons (to be discussed later).

This paper has the following structure. In Sect. \ref{Sect. 2}, the
diffuse-interface model (DIM) is formulated. In Sect. \ref{Sect. 3}, the DIM
is reduced to several simpler sets of equations, depending on the parameters
of the fluid under consideration. In Sect. \ref{Sect. 4}, the properties of
the asymptotic sets are examined, and Sect. \ref{Sect. 5} outlines how the
present results can be made more comprehensive and accurate.

\section{Formulation\label{Sect. 2}}

Consider a flow of a non-ideal fluid characterized by its density
$\rho(\mathbf{r},t)$, velocity $\mathbf{v}(\mathbf{r},t)$, pressure
$p(\mathbf{r},t)$, and temperature $T(\mathbf{r},t)$, where $\mathbf{r}$ is
the position vector and $t$, the time. Let the equation of state be of the van
der Waals type, i.e.,%
\begin{equation}
p=\frac{RT\rho}{1-b\rho}-a\rho^{2}, \label{2.1}%
\end{equation}
where $R$ is the specific gas constant, and $a$ and $b$ are the van der Waals parameters.

There exist several versions of the DIM, which have been applied to numerous
physically-important problems
\cite{HohenbergHalperin77,JasnowVinals96,LowengrubTruskinovsky98,VladimirovaMalagoliMauri99,PismenPomeau00,ThieleMadrugaFrastia07,DingSpelt07,MadrugaThiele09,YueZhouFeng10,YueFeng11,SibleyNoldSavvaKalliadasis13a,SibleyNoldSavvaKalliadasis13b,MagalettiMarinoCasciola15,MagalettiGalloMarinoCasciola16,KusumaatmajaHemingwayFielding16,FakhariBolster17,GalloMagalettiCasciola18,BorciaBorciaBestehornVarlamovaHoefnerReif19,GalloMagalettiCoccoCasciola20,GelissenVandergeldBaltussenKuerten20,Benilov20a}%
. More comprehensive versions (e.g., Refs.
\cite{AndersonMcFaddenWheeler98,ThieleMadrugaFrastia07}) are applicable to
multi-component fluids with variable temperature, simpler ones apply either to
\emph{single}-component \emph{isothermal} fluids (e.g., Ref.
\cite{PismenPomeau00}) or single-component isothermal and
\emph{incompressible} fluids (e.g., Refs.
\cite{JasnowVinals96,DingSpelt07,MadrugaThiele09}).

In the present paper, the non-isothermal compressible DIM for a
single-component fluid will be used, in the form suggested in Ref.
\cite{AndersonMcFaddenWheeler98}.

The governing equations of the version of the DIM suggested in Ref.
\cite{AndersonMcFaddenWheeler98} are%
\begin{equation}
\frac{\partial\rho}{\partial t}+\boldsymbol{\boldsymbol{\nabla}}\cdot\left(
\rho\mathbf{v}\right)  =0, \label{2.2}%
\end{equation}%
\begin{equation}
\frac{\partial\mathbf{v}}{\partial t}+\left(  \mathbf{v}\cdot
\boldsymbol{\boldsymbol{\nabla}}\right)  \mathbf{v}+\frac{1}{\rho
}\boldsymbol{\boldsymbol{\nabla}}\cdot\left(  \mathbf{I}p-\boldsymbol{\Pi
}\right)  =K\boldsymbol{\boldsymbol{\nabla}}\nabla^{2}\rho, \label{2.3}%
\end{equation}%
\begin{multline}
\rho c_{V}\left(  \frac{\partial T}{\partial t}+\mathbf{v}\cdot
\boldsymbol{\boldsymbol{\nabla}}T\right)  +\left(  p+a\rho^{2}\right)
\boldsymbol{\boldsymbol{\nabla}}\cdot\mathbf{v}\\
-\boldsymbol{\Pi}:\boldsymbol{\boldsymbol{\nabla}}\mathbf{v}%
=\boldsymbol{\boldsymbol{\nabla}}\cdot\left(  \kappa
\boldsymbol{\boldsymbol{\nabla}}T\right)  , \label{2.4}%
\end{multline}
where $\mathbf{I}$ is the identity matrix, the viscous stress tensor is%
\begin{equation}
\boldsymbol{\Pi}=\mu_{s}\left[  \boldsymbol{\boldsymbol{\nabla}}%
\mathbf{v}+\left(  \boldsymbol{\boldsymbol{\nabla}}\mathbf{v}\right)
^{T}-\frac{2}{3}\mathbf{I}\left(  \boldsymbol{\boldsymbol{\nabla}}%
\cdot\mathbf{v}\right)  \right]  +\mu_{b}\mathbf{I}\left(
\boldsymbol{\boldsymbol{\nabla}}\cdot\mathbf{v}\right)  , \label{2.5}%
\end{equation}
$\mu_{s}$ ($\mu_{b}$) is the shear (bulk) viscosity, $c_{V}$ is the specific
heat capacity, and $\kappa$ is the thermal conductivity, and the right-hand
side of (\ref{2.3}) represents the so-called Korteweg stress ($K$ is a
fluid-specific constant). Note that $\mu_{s}$, $\mu_{b}$, $c_{V}$, and
$\kappa$ are fluid-specific functions of $\rho$ and $T$.

Let the fluid be enclosed in a container (mathematically speaking, domain)
$\mathcal{D}$, so that%
\begin{equation}
\mathbf{v}=\mathbf{0}\qquad\text{at}\qquad\mathbf{r}\in\partial\mathcal{D},
\label{2.6}%
\end{equation}
where $\partial\mathcal{D}$ is the container's walls (domain's boundary).
Another boundary condition should be imposed on $T$; assuming for simplicity
that the walls are insulated, let%
\begin{equation}
\mathbf{n}\cdot\boldsymbol{\boldsymbol{\nabla}}T=0\qquad\text{at}%
\qquad\mathbf{r}\in\partial\mathcal{D}, \label{2.7}%
\end{equation}
where $\mathbf{n}$ is a normal to $\partial\mathcal{D}$.

Several versions of the boundary condition for $\rho$ exist in the literature
\cite{Seppecher96,PismenPomeau00,SoucekHeidaMalek20}. In this work, the
simplest one is used,%
\begin{equation}
\mathbf{n}\cdot\boldsymbol{\boldsymbol{\nabla}}\rho=0\qquad\text{at}%
\qquad\mathbf{r}\in\partial\mathcal{D}, \label{2.8}%
\end{equation}
which is a particular case of the condition derived in Ref. \cite{Seppecher96}.

Generally, little in the analysis to come depends on the specific form of the
boundary conditions (\ref{2.7})-(\ref{2.8}). They are mostly needed for
numerical simulations reported in Sect. \ref{Sect. 4}.

\section{Simplified models\label{Sect. 3}}

\subsection{Nondimensionalization}

Assuming that the pressure gradient across the interface is balanced by the
Korteweg stress, one can deduce that the spatial scale of interfacial dynamics
is%
\[
\bar{r}=\left(  \frac{K}{a}\right)  ^{1/2}.
\]
Introduce also a velocity scale $\bar{v}$ (so that the time scale is $\bar
{r}/\bar{v}$), a characteristic temperature $\bar{T}$, and the density scale
$b^{-1}$.

The following nondimensional variables will be used:%
\[
\mathbf{r}_{nd}=\frac{\mathbf{r}}{\bar{r}},\qquad t_{nd}=\frac{\bar{v}}%
{\bar{r}}t,
\]%
\[
\rho_{nd}=b\rho,\qquad\mathbf{v}_{nd}=\frac{\mathbf{v}}{\bar{v}},\qquad
T_{nd}=\frac{T}{\bar{T}}.
\]
It is convenient to also introduce the nondimensional versions of the fluid
parameters. Assume for simplicity that the bulk and shear viscosities are of
the same order (say, $\bar{\mu}$), and denote the other two scales by
$\bar{\kappa}$ and $\bar{c}_{V}$, so that%
\[
\left(  \mu_{s}\right)  _{nd}=\frac{\mu_{s}}{\bar{\mu}},\qquad\left(  \mu
_{b}\right)  _{nd}=\frac{\mu_{b}}{\bar{\mu}},
\]%
\[
\kappa_{nd}=\frac{\kappa}{\bar{\kappa}},\qquad\left(  c_{V}\right)
_{nd}=\frac{c_{V}}{\bar{c}_{V}},
\]
and the nondimensional viscous stress is%
\[
\boldsymbol{\Pi}_{nd}=\frac{\bar{r}}{\bar{\mu}\bar{v}}\boldsymbol{\Pi}.
\]
In the most general situation, the viscous stress, the Korteweg stress, and
the pressure gradient in Eq. (\ref{2.3}) are all of the same order, which
implies%
\[
\bar{v}=\frac{a\bar{r}}{\bar{\mu}b^{2}}.
\]
Physically, this scale characterizes the disbalance between the Korteweg
stress and pressure gradient (typically arising if the interface is curved);
most importantly, it has nothing to do with the global flow.

Rewriting Eqs. (\ref{2.1})-(\ref{2.4}) in terms of the nondimensional
variables and omitting the subscript $_{nd}$, one obtains%
\begin{equation}
\frac{\partial\rho}{\partial t}+\boldsymbol{\boldsymbol{\nabla}}\cdot\left(
\rho\mathbf{v}\right)  =0, \label{3.1}%
\end{equation}%
\begin{multline}
\fbox{$\alpha$}\left[  \frac{\partial\mathbf{v}}{\partial t}+\left(
\mathbf{v}\cdot\boldsymbol{\boldsymbol{\nabla}}\right)  \mathbf{v}\right] \\
+\frac{1}{\rho}\boldsymbol{\boldsymbol{\nabla}}\cdot\left[  \mathbf{I}\left(
\frac{\fbox{$\tau$}T\rho}{1-\rho}-\rho^{2}\right)  -\boldsymbol{\Pi}\right]
=\boldsymbol{\boldsymbol{\nabla}}\nabla^{2}\rho, \label{3.2}%
\end{multline}%
\begin{multline}
\fbox{$\alpha\gamma$}\rho c_{V}\left(  \frac{\partial T}{\partial
t}+\mathbf{v}\cdot\boldsymbol{\boldsymbol{\nabla}}T\right) \\
+\fbox{$\beta$}\left(  \frac{\fbox{$\tau$}T\rho}{1-\rho}%
\boldsymbol{\boldsymbol{\nabla}}\cdot\mathbf{v}-\boldsymbol{\Pi}%
:\boldsymbol{\boldsymbol{\nabla}}\mathbf{v}\right)
=\boldsymbol{\boldsymbol{\nabla}}\cdot\left(  \kappa
\boldsymbol{\boldsymbol{\nabla}}T\right)  , \label{3.3}%
\end{multline}
where%
\begin{equation}
\alpha=\frac{K}{\bar{\mu}^{2}b^{3}},\qquad\beta=\frac{aK}{\bar{\mu}\bar
{\kappa}\bar{T}b^{4}}, \label{3.4}%
\end{equation}%
\begin{equation}
\gamma=\frac{\bar{c}_{V}\bar{\mu}}{\bar{\kappa}},\qquad\tau=\frac{R\bar{T}%
b}{a}. \label{3.5}%
\end{equation}
Judging by the positions of $\alpha$ and $\beta$ in Eqs. (\ref{3.3}%
)-(\ref{3.4}), $\alpha$ is the Reynolds number and $\beta$, an `isothermality
parameter'. The latter controls the production of heat due to compressibility
and viscosity, i.e., if $\beta\ll1$, the flow is close to isothermal. In turn,
$\gamma$ is the Prandtl number and $\tau$ is the nondimensional temperature.

It should be emphasized that $\alpha$ and $\beta$ are `microscopic'
parameters. They characterize the flow at the \emph{interfacial} scale and
they do not depend on either the \emph{global} Reynolds number (based on, say,
the droplet's size) nor on whether or not the flow is isothermal
\emph{globally}.

The nondimensional expression for $\boldsymbol{\Pi}$ and the nondimensional
boundary conditions will not be presented, as they look exactly as their
dimensional counterparts (\ref{2.5})-(\ref{2.8}).

\subsection{Asymptotic estimates}

A lot of valuable information can be extracted by estimating the
nondimensional parameters (\ref{3.4})-(\ref{3.5}) for the four liquids for
which discrepancies between experimental and theoretical results have been
reported. To do so, one needs the parameters of these fluids -- all of which,
except $K$, have been found in Ref. \cite{HaynesLideBruno17} and collated, for
the reader's convenience, in Appendix \ref{Appendix A}). $K$, in turn, was
calculated by relating it to the surface tension (see Appendix
\ref{Appendix B}). The temperature scale was set to $\bar{T}=25^{\circ
}\mathrm{C}$, which is regarded in Ref. \cite{HaynesLideBruno17} as the
\textquotedblleft normal temperature\textquotedblright\ and is also close to
the \textquotedblleft room temperature\textquotedblright\ at which experiments
are normally conducted.

At this temperature, the viscosity, specific heat, and thermal conductivity of
the liquid phase of all fluids considered exceed those of the vapor phase by
several orders of magnitude. Thus, the variations of these parameters across
the interface are approximately equal to the liquid values -- which were thus
used to estimate the nondimensional parameters involved.

The estimated values of $\alpha$, $\beta$, $\alpha\gamma$, and $\tau$ are
presented in Table \ref{table1}. The following conclusions can be drawn:

\begin{table*}
\centering
\begin{ruledtabular}
\begin{tabular}{|m{2.7cm}|m{2.5cm}|m{2.5cm}|m{2.5cm}|m{2.5cm}|}
\hspace{9mm}Fluid & \hspace{6mm}$\alpha$ & \hspace{5.5mm} $\beta$ & \hspace{4mm} $\alpha\gamma$ & \hspace{5.5mm} $\tau$ \\[3pt]
\hline
\hspace{0mm} ethylene glycol & \hspace{1mm} $5\times10^{-4}$ & \hspace{3mm} $0.033$  & \hspace{3mm} $0.073$ & \hspace{3mm} $0.123$ \\[3pt]
\hline
\hspace{0mm} glycerol & \hspace{1.5mm} $2\times10^{-7}$ & \hspace{1.5mm} $6\times10^{-4}$ & \hspace{1.5mm} $2\times10^{-3}$ & \hspace{3mm} $0.104$ \\[3pt]
\hline
\hspace{0mm} mercury & \hspace{3mm} $2.63$ & \hspace{3mm} $0.390$ & \hspace{3mm} $0.066$ & \hspace{3mm} $0.050$ \\[3pt]
\hline
\hspace{0mm} water & \hspace{3mm} $0.143$ & \hspace{3mm} $0.711$ & \hspace{3mm} $0.880$ & \hspace{3mm} $0.137$ \\[3pt]
\end{tabular}
\end{ruledtabular}
\caption{The nondimensional parameters (\ref{3.4})-(\ref{3.5}) for fluids under consideration.}
\label{table1}
\end{table*}

\begin{enumerate}
\item The assumption of small Reynolds number, $\alpha\ll1$, does not hold for mercury.

\item The isothermality assumption, $\beta\ll1$, does not hold for mercury,
and even less so, for water.

\item On a less important note, $\tau$ seems to be moderately small for all
four fluids. This impression is misleading, however, as the values of $\tau$
in Table \ref{table1} are comparable to the maximum of this parameter,
$\tau=8/27$ (corresponding to the critical temperature of the van der Waals fluid).
\end{enumerate}

Thus, it comes as no surprise that all of the existing theories of contact
lines fail for mercury and water -- but their failure for $\text{glycerol and
}$ethylene glycol$\text{ must be caused by different reasons. For example, }%
$the discrepancy associated with the latter pair of fluids might be due to
chemical inhomogeneity of the substrate\footnote{The authors of Ref.
\cite{KimLeeKang02} where glycerol and ethylene glycol were examined
specifically state that the mean roughness of the substrate was very low
($1.3~\mathrm{nm}$), but they do \emph{not} mention that the substrate has
been chemically cleaned.}, as inhomogeneities are known to dramatically affect
the dynamics of contact lines \cite{SavvaKalliadasis13}.

In principle, there could be additional reasons for the failure of the
existing models for the four fluids at issue -- but, in case of mercury and
water, these reasons must be sought using non-isothermal models.

\subsection{Asymptotic equations}

Depending on the fluid under consideration, the exact governing equations can
be reduced to a simpler asymptotic set. Three of these will be presented: for
mercury (Set 1), water (Set 2), and glycerol and ethylene glycol (Set 3).

To obtain Set 1, assume%
\[
\alpha\sim1,\qquad\beta\sim1,\qquad\alpha\gamma\ll1,
\]
and omit the terms involving $\alpha\gamma$ from the governing equations. The
density and momentum equations (\ref{3.1})-(\ref{3.2}) remain the same,
whereas Eq. (\ref{3.3}) becomes%
\begin{equation}
\beta\left(  \frac{\tau T\rho}{1-\rho}\boldsymbol{\boldsymbol{\nabla}}%
\cdot\mathbf{v}-\boldsymbol{\Pi}:\boldsymbol{\boldsymbol{\nabla}}%
\mathbf{v}\right)  -\boldsymbol{\boldsymbol{\nabla}}\cdot\left(
\kappa\boldsymbol{\boldsymbol{\nabla}}T\right)  =0. \label{3.6}%
\end{equation}
With the time derivative omitted from this equation, $T$ is `enslaved' by
(instantly adjusts to) the heat production due to compressibility and viscosity.

Given an initial condition for $\rho$ and $\mathbf{v}$, Eqs. (\ref{3.1}%
)-(\ref{3.2}), (\ref{3.6}), expression (\ref{2.5}) for $\boldsymbol{\Pi}$, and
the boundary conditions (\ref{2.6})-(\ref{2.8}) fully determine $\rho
(\mathbf{r},t)$, $\mathbf{v}(\mathbf{r},t)$, $T(\mathbf{r},t)$.

To obtain Set 2, let%
\begin{equation}
\alpha\ll1,\qquad\beta\sim1,\qquad\alpha\gamma\sim1, \label{3.7}%
\end{equation}
and omit the terms involving $\alpha$. The density equation (\ref{3.1}) and
that for the temperature (\ref{3.3}) remain the same, whereas Eq. (\ref{3.2})
becomes%
\begin{equation}
\frac{1}{\rho}\boldsymbol{\boldsymbol{\nabla}}\cdot\left[  \mathbf{I}\left(
\frac{T\rho}{1-\rho}-\rho^{2}\right)  -\boldsymbol{\Pi}\right]
=\boldsymbol{\boldsymbol{\nabla}}\nabla^{2}\rho. \label{3.8}%
\end{equation}
Eqs. (\ref{3.1}), (\ref{3.8}), (\ref{3.3}), and (\ref{2.5}), and the boundary
conditions (\ref{2.6})-(\ref{2.8}) form a full set. This time, the velocity
does not require an initial condition, as it is `enslaved' by $\rho$ and $T$
through Eq. (\ref{3.8}) and boundary condition (\ref{2.6}).

To obtain Set 3, assume%
\[
\alpha\ll1,\qquad\beta\ll1,\qquad\alpha\gamma\ll1.
\]
The density equation (\ref{3.1}) remains as is, the velocity equation is the
same as (\ref{3.8}), whereas (\ref{3.3}) becomes%
\[
\boldsymbol{\boldsymbol{\nabla}}\cdot\left(  \kappa
\boldsymbol{\boldsymbol{\nabla}}T\right)  =0.
\]
This equation and the boundary condition (\ref{2.7}) imply that%
\begin{equation}
T=T(t). \label{3.9}%
\end{equation}
To determine $T(t)$, one needs to return to the exact equation (\ref{3.3}),
integrate it over the domain $\mathcal{D}$ and take into account the boundary
condition (\ref{2.7}) -- so that the leading-order term disappears, resulting%
\begin{equation}
c_{V}M\frac{\mathrm{d}T}{\mathrm{d}t}+T\int_{\mathcal{D}}\frac{\rho}{1-\rho
}\boldsymbol{\boldsymbol{\nabla}}\cdot\mathbf{v}\,\mathrm{d}^{3}%
\mathbf{r}-\int_{\mathcal{D}}\boldsymbol{\Pi}:\boldsymbol{\boldsymbol{\nabla}%
}\mathbf{v}\,\mathrm{d}^{3}\mathbf{r}=0, \label{3.10}%
\end{equation}
where%
\[
M=\int_{\mathcal{D}}\rho\,\mathrm{d}^{3}\mathbf{r}%
\]
is constant due to the mass conservation law.

Eqs. (\ref{3.1}), (\ref{3.8}), (\ref{3.10}), and (\ref{2.5}), and the boundary
conditions (\ref{2.6}), (\ref{2.8}) form a full set. The initial condition for
$T$ should not depend on the spatial variables, as initial variations of $T$
(if any) are implied to rapidly even out, so that the flow almost instantly
becomes isothermal.

In most applications, the container is so large that $M\gg1$. In this case,
Eq. (\ref{3.10}) yields $\mathrm{d}T/\mathrm{d}t\approx0$ -- hence, in the
other equations, $T$ can be treated as a known constant determined by the
initial condition. The resulting model is mathematically equivalent to the one
examined in Ref. \cite{PismenPomeau00}.

\section{Properties of the asymptotic models \label{Sect. 4}}

Given that it is nearly impossible to separate phase transition from
hydrodynamic motion, it is vital that the derived asymptotic models satisfy
the fundamental requirements of thermodynamics: firstly, they should comply
with the Maxwell construction and, secondly, predict the correct threshold of
the instability responsible for phase transitions. In what follows, both
requirements will be illustrated for the simplest of the sets derived, Set 3.

Consider the one-dimensional reduction of Set 3, i.e., let $v_{1}=v_{2}=0$,
with the rest of the unknowns depending only on $r_{3}$ and $t$. Denoting
$v_{3}=w$ and $r_{3}=z$, and considering for simplicity the large-container
limit, one can reduce Eqs. (\ref{3.1}), (\ref{3.8}), and \ref{2.5}), and the
boundary conditions (\ref{2.6}) and (\ref{2.8}) to%
\begin{equation}
\frac{\partial\rho}{\partial t}+\frac{\partial\left(  \rho w\right)
}{\partial z}=0, \label{4.1}%
\end{equation}%
\begin{equation}
\frac{1}{\rho}\frac{\partial}{\partial z}\left(  \frac{T\rho}{1-\rho}-\rho
^{2}-\eta\frac{\partial w}{\partial z}\right)  =\frac{\partial^{3}\rho
}{\partial z^{3}}, \label{4.2}%
\end{equation}%
\begin{equation}
w=0,\qquad\frac{\partial\rho}{\partial z}=0\qquad\text{at}\qquad z=\pm\frac
{1}{2}Z, \label{4.3}%
\end{equation}
where $\eta=\frac{4}{3}\mu_{s}+\mu_{b}$ and $Z$ is the container size.

It can be readily shown that:

\begin{itemize}
\item Steady solutions -- such that $w=0$, $\rho=\rho(z)$ -- of Eqs.
(\ref{4.1})-(\ref{4.3}) with $Z=\infty$ describe a stationary liquid/vapor
interface in an infinite domain. It can be readily shown that, for these
solutions,%
\[
\lim\limits_{_{z\rightarrow-\infty}}\left(  \frac{T\rho}{1-\rho}-\rho
^{2}\right)  =\lim\limits_{z\rightarrow-\infty}\left(  \frac{T\rho}{1-\rho
}-\rho^{2}\right)  ,
\]%
\begin{multline*}
\lim\limits_{z\rightarrow-\infty}\left[  T\left(  \ln\frac{\rho}{1-\rho}%
+\frac{1}{1-\rho}\right)  -2\rho\right] \\
=\lim\limits_{z\rightarrow\infty}\left[  T\left(  \ln\frac{\rho}{1-\rho}%
+\frac{1}{1-\rho}\right)  -2\rho\right]  ,
\end{multline*}
which is the van der Waals version of the Maxwell construction, according to
which the pressures and the densities of Gibbs free energy of the two phases
must be equal.

\item As shown in Appendix \ref{Appendix C}, a single-phase state
characterized by a pair $\left(  T,\rho\right)  $ and governed by Eqs.
(\ref{4.1})-(\ref{4.3}) in an infinite domain is unstable, if%
\begin{equation}
\frac{T}{\left(  1-\rho\right)  ^{2}}-2\rho<0, \label{4.4}%
\end{equation}
which is the van der Waals version of the thermodynamic instability criterion
\cite{FerzigerKaper72}%
\[
\left(  \dfrac{\partial p}{\partial\rho}\right)  _{T}<0.
\]
Given that this instability triggers off phase transitions, one should hope
that the asymptotic equations describe those well.
\end{itemize}

The boundary-value problem (\ref{4.1})-(\ref{4.3}) was simulated numerically
using the method of lines \cite{Schiesser78}, for various initial conditions
and various examples of the viscosity $\eta(\rho,T)$. As expected, two
patterns of dynamics were observed: the solution would evolve either toward a
single-phase state or a two-phase state.

\begin{figure}
\includegraphics[width=\columnwidth]{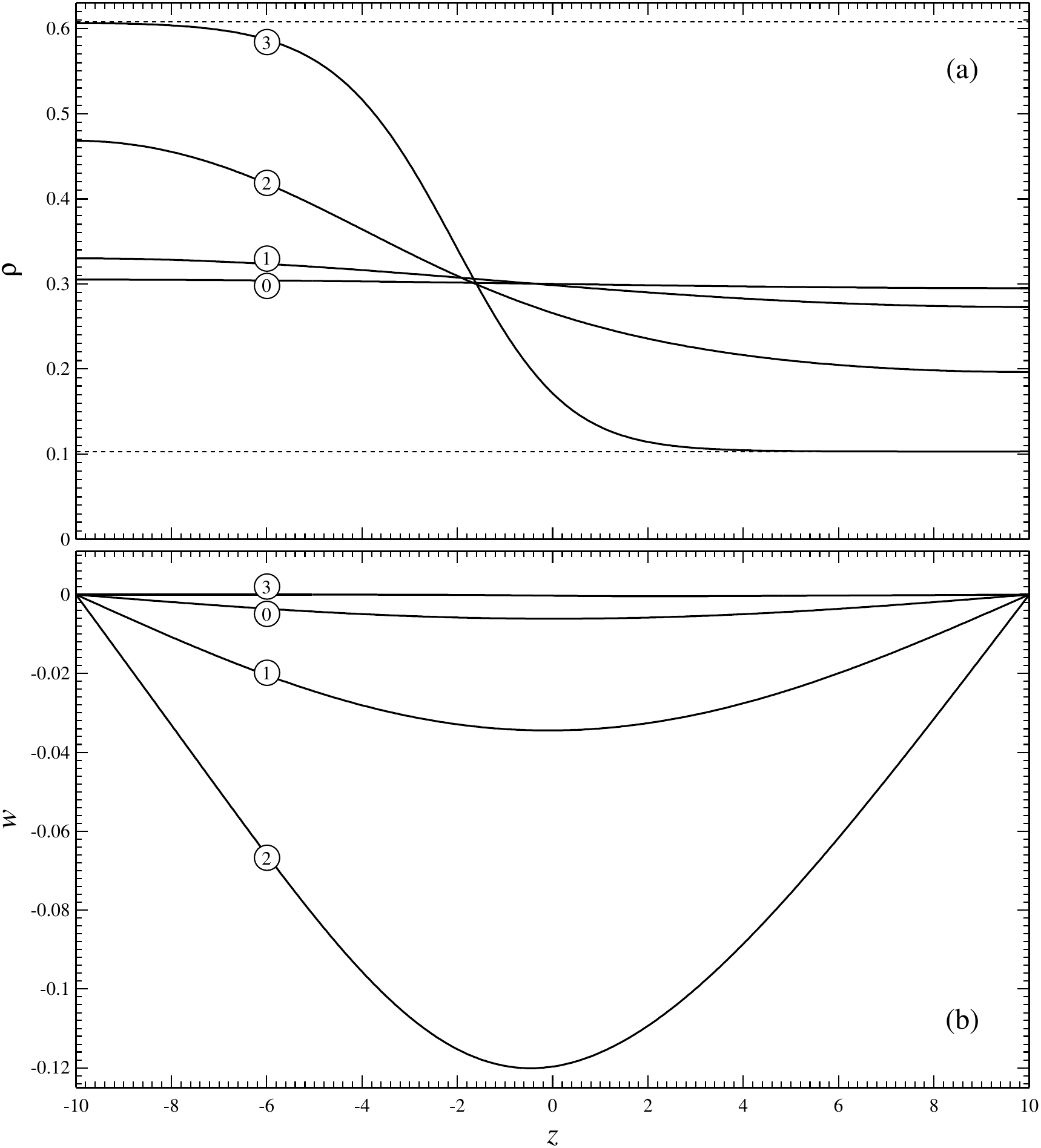}
\caption{The solution of Eqs. (\ref{4.1})-(\ref{4.3}) and (\ref{4.5}), with $Z=20$ and the initial condition (\ref{4.6}). The curves labeled \textquotedblleft0\textquotedblright, \textquotedblleft 1\textquotedblright, \textquotedblleft2\textquotedblright, \textquotedblleft 3\textquotedblright\ correspond to $t=0,~30,~60,~300$, respectively. (a) The density field (the dotted lines show the liquid and vapor densities predicted by the Maxwell construction). (b) The velocity field.}
\label{fig1}
\end{figure}

An example of the latter behavior was computed for the following
(nondimensional) viscosity\footnote{The dependence of $\eta$ on the
temperature can be ignored, as $T$ does not change in time in Set 3. Otherwise
expression (\ref{4.5}) appears to be a good qualitative model of the real
dependence of viscosity of water on its density \cite{LinstromMallard97}.}:
\begin{equation}
\eta=\frac{\rho}{1-\rho}, \label{4.5}%
\end{equation}
for the nondimensional temperature%
\begin{equation}
T=0.25, \label{4.6}%
\end{equation}
and the initial condition%
\begin{equation}
\rho=0.3+0.005\sin\frac{\pi z}{Z}. \label{4.7}%
\end{equation}
Criterion (\ref{4.4}) predicts that steady state (\ref{4.6})-(\ref{4.7}) is
unstable -- which it indeed is, as can be seen in Fig. \ref{fig1}. Evidently,
the solution evolves into the two-phase state described by the Maxwell construction.

\section{Concluding remarks\label{Sect. 5}}

Thus, depending on the fluid under consideration, the diffuse-interface model
can be reduced to one of three possible sets of asymptotic equations. Only one
of the three satisfies the assumptions on which the existing models of contact
lines are based, so no new results should be expected in this case. The other
two asymptotic sets should go beyond the existing models, describing fluids to
which these models do not apply (such as water and mercury).

In addition to the four fluids included in the present paper another four have
been examined (acetone, benzene, ethanol, and methanol). Only for one of these
$\alpha$ is small, and none have isothermal interfaces -- which makes one
wonder whether the failure of these assumptions is an exception or rule. They
seems to hold only for high-viscosity fluids, such as glycerol and ethylene
glycol, as well as (probably) silicone oils which are frequently used in
experiments with contact lines. This hypothesis, however, remains unverified,
as the full set of characteristics of any of silicone oils does not seem to be
available, neither in the literature nor internet.

As this work is only a proof of concept, there are a number of extensions of
the DIM to be considered in the future -- such as introduction of pair
correlations \cite{MaitrejeanAmmarChinestaGrmela12}, non-Newtonian viscosity,
non-Fourier heat conduction, and self-diffusion
\cite{Grmela14,VanPavelkaGrmela17}. Such extensions should be relatively easy
do develop using the so-called GENERIC tool
\cite{GrmelaOttinger97,OettingerGrmela97}, and they should make the asymptotic
models proposed in this paper more comprehensive and accurate. It is also
crucial to give up the van der Waals equation of state and use a realistic
one, describing the fluid under consideration with a sufficient accuracy (this
has already been done for water \cite{Benilov20a}).

\appendix

\section{The parameters of the fluids under consideration\label{Appendix A}}

All of the parameters listed in this appendix have been taken from Ref.
\cite{HaynesLideBruno17}.

The van der Waals constants $a$ and $b$ were calculated using the critical
temperature $T_{c}$ and the critical pressure $p_{c}$, through the formulae
(see Ref. \cite{HaynesLideBruno17})%
\begin{equation}
a=\frac{27R^{2}T_{c}^{2}}{64p_{c}m^{2}},\qquad b=\dfrac{RT_{c}}{8p_{c}m},
\label{A.1}%
\end{equation}
where $m$ is the molar mass. The results, as well as the `source data', are
presented in Table \ref{table2}.

\begin{table*}
\centering
\begin{ruledtabular}
\begin{tabular}{|m{2.5cm}|m{2.5cm}|m{2.5cm}|m{2.5cm}|m{3cm}|m{2.5cm}|}
\hspace{9mm}Fluid & \hspace{1mm} $m~(\mathrm{g~mol}^{-1})$ & \hspace{5mm} $T_{c}~(\mathrm{K})$ & \hspace{3mm} $p_{c}~(\mathrm{MPa})$ & \hspace{1.5mm} $a~(\mathrm{m}^{5}\mathrm{s}^{-2}\mathrm{kg}^{-1})$ & \hspace{2mm} $b~(\mathrm{m}^{3}\mathrm{g}^{-1})$ \\[3pt]
\hline
\hspace{0mm} ethylene glycol & \hspace{6mm} $62.07$ & \hspace{7.5mm} $719$  & \hspace{8mm} $8.1$ & \hspace{8mm} $483.13$ & \hspace{5mm} $1.4863$\\[3pt]
\hline
\hspace{0mm} glycerol & \hspace{6mm} $92.09$ & \hspace{7.5mm} $850$ & \hspace{8mm} $7.6$ & \hspace{8mm} $326.93$ & \hspace{5mm} $1.2622$ \\[3pt]
\hline
\hspace{0mm} mercury & \hspace{4.5mm} $200.59$ & \hspace{6mm} $1764$ &\hspace{4.5mm} $167$ & \hspace{9.5mm} $13.506$ & \hspace{5mm} $0.0547$ \\[3pt]
\hline
\hspace{0mm} water & \hspace{6mm} $18.02$ & \hspace{7.5mm} $647.10$ &\hspace{6mm} $22.06$ & \hspace{6.5mm} $1704.8$ & \hspace{5mm} $1.6918$ \\[3pt]
\end{tabular}
\end{ruledtabular}
\caption{
The molar masses, the critical temperatures and pressures, and the van der Waals parameters [determined by (\ref{A.1})] of the fluids under consideration.}
\label{table2}
\end{table*}

\begin{table*}
\centering
\begin{ruledtabular}
\begin{tabular}{|m{2.5cm}|m{2.5cm}|m{2.5cm}|m{2.5cm}|m{2.5cm}|}
\hspace{9mm}Fluid & \hspace{0mm} $\bar{\mu}~(\mathrm{mPa~s})$ & \hspace{-5.5mm} $\bar{\kappa}~(\mathrm{W~m}^{-1}\mathrm{K}^{-1})$ & \hspace{-6.5mm} $\bar{c}_{p}~(\mathrm{kJ~kg}^{-1}\mathrm{K}^{-1})$ & \hspace{-3mm} $\sigma~(\mathrm{mN~m}^{-1})$ \\[3pt]
\hline
\hspace{0mm} ethylene glycol & \hspace{2mm} $16.06$ & \hspace{2mm} $0.254$  & \hspace{2mm} $2.394$ & \hspace{2mm} $48.02$ \\[3pt]
\hline
\hspace{0mm} glycerol & \hspace{0.5mm} $934$ & \hspace{2mm} $0.285$ & \hspace{2mm} $2.377$ & \hspace{2mm} $62.5$ \\[3pt]
\hline
\hspace{0mm} mercury & \hspace{4mm} $1.526$ & \hspace{2mm} $8.514$ & \hspace{2mm} $0.114$ & \hspace{0.7mm} $485.48$ \\[3pt]
\hline
\hspace{0mm} water & \hspace{4mm} $0.890$ & \hspace{2mm} $0.6062$ & \hspace{2mm} $4.179$ & \hspace{2mm} $72.06$ \\[3pt]
\end{tabular}
\end{ruledtabular}
\caption{The dynamic viscosities, thermal conductivities, specific heat capacities, and surface tensions of the fluids under consideration (all
at $25^{\circ}\mathrm{C}$).}
\label{table3}
\end{table*}

Table \ref{table3}, in turn, presents the dynamic viscosities, thermal
conductivities, specific heat capacities, and surface tensions. Note that Ref.
\cite{HaynesLideBruno17} does not present data on $\bar{c}_{V}$ which was used
for nondimensionalizing the governing equations, so $\bar{c}_{p}$ was used
instead, so that the assumption $\bar{c}_{V}\approx\bar{c}_{p}$ was implied.
Admittedly, it does not hold for gases, but does do for liquids (for water,
for example, $\bar{c}_{V}\approx4.13~\mathrm{kJ~kg}^{-1}\mathrm{K}^{-1}$ and
$\bar{c}_{p}\approx4.18~\mathrm{kJ~kg}^{-1}\mathrm{K}^{-1}$). Besides,
$\bar{c}_{p}$ is used in this paper as a \emph{scale} for $\bar{c}_{V}$, so
its precise value is unimportant.

\section{Deducing $K$ from a liquid's surface tension\label{Appendix B}}

Within the framework of the DIM, the surface tension of a liquid/vapor
interface can be related to the solution of the static one-dimensional
reduction of Eqs. (\ref{2.1})-(\ref{2.5}). Setting, accordingly,
$\partial/\partial t=0$, $\mathbf{v}=\mathbf{0}$, and $\rho=\rho(z)$, one
obtains%
\begin{equation}
\frac{1}{\rho}\left[  \frac{RT}{\left(  1-b\rho\right)  ^{2}}-2a\rho\right]
\frac{\mathrm{d}\rho}{\mathrm{d}z}=K\frac{\mathrm{d}^{3}\rho}{\mathrm{d}z^{3}%
}. \label{B.1}%
\end{equation}
This equation is to be solved in an unbounded domain under the condition%
\begin{equation}
\frac{\mathrm{d}\rho}{\mathrm{d}z}\rightarrow0\qquad\text{as}\qquad
z\rightarrow\pm\infty. \label{B.2}%
\end{equation}
Once the boundary-value problem (\ref{B.1})-(\ref{B.2}) is solved and its
solution $\rho(z)$ is found, the surface tension of liquid/vapor interface is
given by \cite{Mauri13}%
\begin{equation}
\sigma=K\int_{-\infty}^{\infty}\left(  \frac{\mathrm{d}\rho}{\mathrm{d}%
z}\right)  ^{2}\mathrm{d}z. \label{B.3}%
\end{equation}
Now, assume that the real-life value of $\sigma$ has been measured at a
certain temperature $\bar{T}$. To determine $K$ in this case, one should solve
the boundary-value problem (\ref{B.1})-(\ref{B.2}) for $T=\bar{T}$ while
varying $K$ -- until the result computed through (\ref{B.3}) coincides with
the measured $\sigma$. Note that, even though this approach depends on the
choice of $\bar{T}$, the resulting $K$ is supposed to apply to the whole
temperature range between the triple and critical points (as the DIM assumes
that $K$ does not depend on $T$).

Computed with $\bar{T}=25^{\circ}\mathrm{C}$, the values of $K$ for the fluids
under consideration are presented in Table \ref{table4}.

\begin{table}
\centering
\begin{ruledtabular}
\begin{tabular}{|m{2.5cm}|m{2.5cm}|}
\hspace{9mm}Fluid & \hspace{-21mm} $K\times10^{16}~(\mathrm{m}^{7}\mathrm{kg}^{-1}\mathrm{s}^{-2})$ \\[3pt]
\hline
\hspace{0mm} ethylene glycol & \hspace{-9.5mm} $16.06$ \\[3pt]
\hline
\hspace{0mm} glycerol & \hspace{-11mm} $934$ \\[3pt]
\hline
\hspace{0mm} mercury & \hspace{-7.5mm} $1.526$ \\[3pt]
\hline
\hspace{0mm} water & \hspace{-7.5mm} $0.890$ \\[3pt]
\end{tabular}
\end{ruledtabular}
\caption{The Korteweg parameter of the fluids under consideration, computed using the approach described in Appendix \ref{Appendix B}}
\label{table4}
\end{table}

Note that expression (\ref{B.3}) represents the surface tension of a
liquid/\emph{vapor} interface -- whereas the data$\ $in Ref.
\cite{HaynesLideBruno17} are for the liquid/\emph{air }one. However, these
parameters are close: for water at $25^{\circ}\mathrm{C}$, for example, the
former is $\sigma=71.97~\mathrm{mN~m}^{-1}$ \cite{WagnerKretzschmar08} and the
latter is $\sigma=72.06$ $\mathrm{mN~m}^{-1}$ \cite{HaynesLideBruno17}.

\section{Derivation of the instability criterion (\ref{4.4})\label{Appendix C}%
}

Consider a homogeneous state characterized by a density $\bar{\rho}$ and
temperature $\bar{T}$; assume also that the fluid is at rest, $\bar{w}=0$, and
let the solution have the form%
\[
\rho=\bar{\rho}+\tilde{\rho}(t,z),\qquad w=\tilde{w}(t,z),
\]
where the tilded variable represent a small perturbation. Substituting the
above expressions into Eqs. (\ref{4.1})-(\ref{4.2}), then linearizing them and
omitting overbars, one obtains%
\begin{equation}
\frac{\partial\tilde{\rho}}{\partial t}+\rho\frac{\partial\tilde{w}}{\partial
z}=0, \label{C.1}%
\end{equation}%
\begin{equation}
\frac{1}{\rho}\frac{\partial}{\partial z}\left[  \frac{T\tilde{\rho}}{\left(
1-\rho\right)  ^{2}}-2\rho\tilde{\rho}-\eta\frac{\partial\tilde{w}}{\partial
z}\right]  =\frac{\partial^{3}\tilde{\rho}}{\partial z^{3}}, \label{C.2}%
\end{equation}
Only harmonic disturbances will be examined, i.e.%
\begin{equation}
\tilde{\rho}=\hat{\rho}\operatorname{e}^{ikz+\lambda t},\qquad\tilde{w}%
=\hat{w}\operatorname{e}^{ikz+\lambda t}, \label{C.3}%
\end{equation}
where $k$ is the perturbation's wavenumber and $\lambda$, its growth/decay
rate. If, for some $k$, $\operatorname{Re}\lambda>0$, the state characterized
by $\left(  \rho,T\right)  $ is unstable.

Substituting (\ref{C.3}) into (\ref{C.1})-(\ref{C.2}), one obtains%
\[
\lambda\hat{\rho}+i\rho k\hat{w}=0,\qquad\frac{1}{\rho}\left[  \frac
{T\hat{\rho}}{\left(  1-\rho\right)  ^{2}}-2\rho\hat{\rho}-ik\eta\hat
{w}\right]  =-k^{2}\hat{\rho}.
\]
These equations admit a solution for $\hat{\rho}$ and $\hat{w}$ only if%
\[
\lambda=-\frac{\rho}{\eta}\left[  \frac{T}{\left(  1-\rho\right)  ^{2}}%
-2\rho+k^{2}\rho\right]  ,
\]
which shows that a value of $k$ exists such that $\lambda>0$ only subject to
condition (\ref{4.4}).

\bibliography{}

\end{document}